\documentclass[aip,graphicx]{revtex4-1}

\usepackage{graphicx}

\usepackage{mathrsfs}

\usepackage{amsmath,amsthm,amsfonts, latexsym}

\newcommand{\mtx}[2]{\left(\begin{array}{#1}#2\end{array}\right)}

\begin{document}


\title{States that ``look the same" with respect to every basis in a mutually unbiased set} 



\author{Ilya Amburg}
\affiliation{Department of Physics, Williams College, Williamstown, MA 01267, USA}
\author{Roshan Sharma}
\affiliation{Applied Physics and Applied Mathematics Department, Columbia University, New York, NY 10024, USA}
\author{Daniel M. Sussman}
\affiliation{Department of Physics and Astronomy, University of Pennsylvania, \\209 South 33rd St., Philadelphia, PA 19104, USA}
\author{William K. Wootters}
\affiliation{Department of Physics, Williams College, Williamstown, MA 01267, USA}



\begin{abstract}
A complete set of mutually unbiased bases in a Hilbert space of dimension $d$ defines
a set of $d+1$ orthogonal measurements.  
Relative to such a set, we define a {\em MUB-balanced} state
to be a pure state for which the list of probabilities of the $d$ outcomes of any of these measurements
is independent of the choice of measurement, up to permutations.  
In this paper we explicitly construct a MUB-balanced state for each prime power dimension $d$ for which
$d = 3$ (mod 4).  These states have already been constructed by Appleby in unpublished notes,  
but our presentation here is different in that both the expression for the states themselves and the proof of MUB-balancedness are given 
in terms of the discrete Wigner function, rather than the density matrix or state vector.  
The discrete Wigner functions of these states are ``rotationally symmetric'' in a sense 
roughly analogous to the rotational symmetry of the energy eigenstates of a harmonic oscillator in the continuous two-dimensional phase
space.
Upon converting the Wigner function to a density matrix, we find that the states are expressible as real state vectors in the standard basis.
We observe numerically that when $d$ is large (and not a power of 3), a histogram of the components of such a state vector appears to 
form a semicircular distribution.
\end{abstract}


\maketitle 

\section{Introduction}  \label{secone}

Consider any energy eigenstate of a simple harmonic oscillator.  Given a system in such a state, suppose we measure
the observable $(\cos\theta) \hat{q} + (\sin\theta) \hat{p}$, where $\theta$ is real and $\hat{q}$
and $\hat{p}$ are the position and momentum operators scaled so that the Hamiltonian is proportional to $\hat{q}^2 + \hat{p}^2$.  
We find that the resulting probability distribution is independent
of $\theta$.  This property is closely related to a symmetry of the state's 
Wigner function\cite{Wigner,Wignerreview}: again with suitable scaling of the position and momentum axes, the Wigner function of any harmonic
oscillator eigenstate is circularly symmetric around the origin of the two-dimensional phase space. 

Our aim in this paper is to find state vectors in a finite-dimensional Hilbert space that have properties analogous to the above
properties of harmonic oscillator eigenstates.  In finite dimension, the closest analog of the set of measurements of the 
form $(\cos\theta) \hat{q} + (\sin\theta) \hat{p}$ is the set of measurements defined by a complete set of mutually unbiased bases (MUBs).\cite{WoottersWigner}
Two orthonormal bases
in dimension $d$ are called mutually unbiased if the inner product between any vector in one of the bases and any vector in the other
basis has magnitude $1/\sqrt{d}$.  That is, $|\langle b^{(1)}_j|b^{(2)}_k\rangle| = 1/\sqrt{d}$ for all $j$ and $k$, where $|b^{(\mu)}_j\rangle$ is the 
$j^{th}$ vector in the $\mu^{th}$ basis.  Such bases are ``as different as possible'' from each other.\cite{Schwinger, Ivanovic, WoottersFields}  It is known that 
in any dimension $d$, one can find at most $d+1$ bases that are pairwise mutually unbiased, and this bound can be achieved when $d$ is a power of a 
prime.\cite{WoottersFields, Bandyopadhyay, MUBreview}  It is not known whether the bound can be achieved for any other value of $d$, though there is strong evidence
against this possibility for the smallest such value, $d=6$.\cite{sixevidence7, sixevidence6, sixevidence5, Skinner, sixevidence4, sixevidence3, sixevidence2, sixevidence1}

To identify an analog of the first of the above properties of harmonic oscillator eigenstates, we consider a Hilbert space in which
a complete set of $d+1$ MUBs exists.  Let the bases be labeled by $\mu = 0, \ldots, d$.  Relative to such a set, we will say that a pure state $|\psi\rangle$ in the 
Hilbert space is ``MUB-balanced'' if the list of probabilities, $(|\langle \psi|b^{(\mu)}_1\rangle|^2, \ldots, |\langle \psi|b^{(\mu)}_d\rangle|^2)$,
is independent of $\mu$, up to permutations.  That is, given a MUB-balanced state, if one were to perform on the state the measurement corresponding to one of the 
MUBs, it would not be possible to tell, just from the set of probabilities of the outcomes, which measurement
was being performed.  

The interest in such states comes partly from a desire to understand the relation between discrete and continuous quantum mechanics.  
We will see that there are intriguing differences between the two cases.  On a more practical note, complete sets of mutually unbiased bases have
been used in the construction of quantum key distribution and secret-sharing schemes\cite{QKD9, QKD8, QKD7, QKD6, QKD5, QKD4, QKD3, QKD2}, for which MUB-balanced states could play a useful conceptual role.
In thinking about intercept-resend eavesdropping attacks, for example, Brierley has noted that when the legitimate participants in a quantum cryptographic scheme are using a complete set of mutually unbiased bases, there is no orthogonal 
measurement an eavesdropper 
could use that is ``halfway'' between all
of these bases.\cite{QKD3}  However, if there exists a MUB-balanced state, it could be used to define a {\em non-orthogonal} measurement that would be related in essentially the same way to each of the mutually unbiased
bases (see Section \ref{secfour} below) and would in this sense be an analog of the ``halfway-between'' measurement in the BB84 scheme.\cite{BB84}
The concept of a MUB-balanced state is also closely related to the concept of a ``minimum uncertainty state,'' which 
has been studied in a number of earlier papers\cite{WoottersSussman, Sussmanthesis, Fuchs, Applebypublished, Galvao}
and has been connected to symmetric measurements\cite{Fuchs} and quantum
random-access codes.\cite{Galvao}  (See Section \ref{secsix} below for a discussion
of the relation between ``MUB-balanced'' and ``minimum uncertainty.'')  

Finally, there is always a certain mathematical interest in
finding states that, according to parameter-counting arguments, have no right to exist.  In this case, one can reasonably 
argue that, by asking that a state vector be MUB-balanced, we are 
imposing $d(d-1)$ constraints on the vector: for the first measurement, the $d-1$ independent probabilities are unconstrained at first, but then for each of the
other $d$ measurements, the $d-1$ probabilities have to match those of the first measurement.  (The permutation freedom is discrete
and does not change the number of parameters.)  But a pure state is specified by only $2(d-1)$ real numbers.  So we seem to be over-constraining
the state by a factor of $d/2$.  Of course the equations that have to be satisfied are nonlinear; so we cannot assume that the parameter-counting argument
is reliable.  Still, if such a state exists, this fact tells us that there is something special about the structure of 
the complete set of MUBs, that allows the state to ``beat the odds.''

For the case $d=2$, one can see immediately that several MUB-balanced states exist.  In that case a convenient complete set of
mutually unbiased bases consists of the bases of eigenstates of the three Pauli operators $X$, $Y$, and $Z$.  A MUB-balanced state
would be any pure state that, on the Bloch sphere, makes equal angles with the $x$, $y$, and $z$ axes.  Of course it becomes much harder
to imagine a MUB-balanced state as the dimension increases.  In this paper we explicitly construct a MUB-balanced state for each prime power dimension
$d$ for which $d=3$ (mod 4).  Moreover, once one such state has been identified, it can be used to generate several others, as we will see
in Section \ref{secfour}.  

We could pose the question of the existence of a MUB-balanced state for any dimension $d$ for which a complete set of MUBs exists.
So we could consider any value of $d$ that is a power of a prime.  The case $d=2^n$ has in effect already been treated 
in earlier papers.\cite{QKD6, Gow, WoottersSussman, Kern, Seyfarth}  Though those papers 
were addressing slightly different questions---the cyclic generation of mutually unbiased bases or the existence of minimum uncertainty states---the arguments given there show directly that MUB-balanced states exist for every $d=2^n$.  
The case we consider here, with $d$ being a prime power equivalent to 3 (mod 4), has in fact also been considered before, in unpublished notes by Appleby\cite{Applebyunpublished}, again
addressing the closely related concept of a minimum uncertainty state.  It follows from Appleby's argument---which is similar to a more specialized argument by Sussman\cite{Sussmanthesis}---that the 
minimum uncertainty state he constructs for any such dimension is also a MUB-balanced state.  
(See also the new paper by Appleby, Bengtsson and Dang.\cite{ABD})
However, the proof we present here is self-contained and is different from Appleby's, though it
is certainly related.  One unusual feature of our proof is that it is based entirely on the {\em discrete Wigner function} of the 
special state (see below) rather than its state vector or density matrix.  
It turns out that our argument does not work at all for $d=1$ (mod 4), and it appears to be an open question whether a
MUB-balanced state exists in any of those cases.

Just as a harmonic oscillator eigenstate has a circularly symmetric Wigner function, the states we identify as MUB-balanced
have a kind of circular symmetry in a discrete phase space.  Here we take the discrete phase space to consist of the 
elements of ${\mathbb F}_d^2$, that is, the two-dimensional vector space over the finite field with $d$ elements.  The phase space can be pictured
as a $d \times d$ array of points, labeled by two coordinates $q$ and $p$ that take values in ${\mathbb F}_d$.  The discrete Wigner function is a representation of a quantum state
as a real function on this phase space.  For our special MUB-balanced state, the Wigner function is constant on each ``circle,'' defined as the set of solutions $(q,p)$
of an equation of the form $q^2 + p^2 = c$ with $c \ne 0$.  It is in this sense that the Wigner function is circularly symmetric.  
(A different analog of an energy eigenstate state of a harmonic oscillator has been investigated by other authors.\cite{Klimov2})

Of particular interest for our purpose is the connection between the discrete Wigner function and a complete set of 
mutually unbiased bases.  As we will discuss in greater detail in the following section, in the discrete phase space we can speak of
``lines" and ``parallel lines,'' each line consisting of exactly $d$ points.  There are $d+1$ possible slopes of a line, and for each value of the 
slope, the $d^2$ points of the phase space can be partitioned into $d$ parallel lines having that slope.  We call such a set of parallel lines
a ``striation.''
Moreover, each striation is associated with one of the bases in a complete set of MUBs, in the following sense: given a quantum state
represented by its Wigner function, if we sum the Wigner function over the $d$ lines of a striation, we obtain the probabilities of the outcomes
of the orthogonal measurement associated with that striation.  Thus a state is MUB-balanced (relative to the set of MUBs associated 
with the discrete Wigner function) if and only if its Wigner function yields the same
list of numbers (up to permutation) when summed over any striation.  The notion of circular symmetry enters the argument as a way of achieving this invariance,
as we explain in Section \ref{secfour}.  
We will see that the role of circular symmetry is somewhat
more subtle than in the case of a continuous phase space.  

For the class of $d$'s we consider, and for the representation of MUBs we use, we find that the MUB-balanced state we identify is representable as a real vector in the standard basis.
Numerical evaluation of the components of this vector reveals an intriguing feature: for large $d$, a histogram of the values of the components typically
appears to form a semicircular distribution (though not when $d$ is a power of 3).  That this appearance reflects a genuine limiting behavior has in fact now been 
proved, and in greater generality, in a recent paper by Katz.\cite{Katz}

Though it has been proved that for each of the dimensions $2$ through 5 there is only one complete set of MUBs up to unitary equivalence,\cite{BSTW, BWB} it is known that 
in many higher dimensions unitarily inequivalent complete sets exist.\cite{Kantor}  In the following sections, we consider only a specific class of MUBs associated with a discrete phase space as described above.  However, our
results directly imply the existence of MUB-balanced states relative to any equivalent set of MUBs.  (Our results say nothing about unitarily inequivalent sets of MUBs.)  
The question of whether and in what sense
the observed semicircular distribution carries over to such equivalent sets of MUBs is more subtle and we do not explore that question here.



We begin in the following section by defining the Wigner function and explaining more fully its relation to mutually unbiased bases.  
We conclude that section by writing down an expression for the Wigner function of our special state.  In Section \ref{secthree} we prove that this expression
does indeed define a pure quantum state, and in Section \ref{secfour} we prove that the state is MUB-balanced.  Then in Section \ref{secfive} we write down the density 
matrix of the state and present a histogram of the values of the components
of the state vector for a typical large value of $d$.  One sees there the approximate semicircular distribution mentioned above.  The final section
summarizes our results and makes a connection with minimum uncertainty states.

\section{The discrete Wigner function} \label{sectwo}

Again, the discrete Wigner function is a representation of a quantum state as a real function on discrete phase space.  The state
could be pure or mixed---the Wigner function contains exactly the information normally expressed in the density matrix.  (For example, multiplying 
a state vector by an 
overall phase factor does not change its Wigner function.)
For our work here, it will be convenient to use the term ``Wigner function'' somewhat more broadly, to refer to a
representation in phase space of an arbitrary Hermitian operator on the $d$-dimensional Hilbert space, not just a density operator.

Several different discrete Wigner functions have been defined in the literature\cite{Gibbons, discreteWigreview} (see also
the references cited in those two papers).
In this paper we use the version of the Wigner function that seems to have first appeared in papers by Klimov and Mu\~noz\cite{Klimov1} and by Vourdas\cite{Vourdas}; it is a particularly simple and natural case 
of a broad class of generalized discrete Wigner functions based on finite fields.\cite{Gibbons}  For odd prime dimensions, this Wigner function is equivalent to discrete Wigner functions\cite{WoottersWigner,Gross1} that have been shown to be especially useful for the analysis of 
quantum computing.\cite{Gross1, Gross2, Emerson1, Emerson2}  

Both our discrete Wigner function and our later arguments are couched in terms of finite fields, so we begin by recalling a few basic facts about such fields.\cite{Lidl}  First, there exists a field with $d$ elements if and only if
$d$ is a power of a prime, and for any such value there is only one field up to isomorphism.  We are calling it ${\mathbb F}_d$.  When $d$ is a prime number, ${\mathbb F}_d$ is the same as ${\mathbb Z}_d$, that is, the
set $\{0, 1, \ldots, d-1\}$ with addition and multiplication mod $d$, but there is no such equivalence for other prime powers.  For $d=r^n$ with $r$ prime, the elements of ${\mathbb F}_d$ can be written as
\begin{equation}  \label{vectorspace}
x = x_0 + x_1\beta + x_2 \beta^2 + \cdots + x_{n-1}\beta^{n-1},
\end{equation}
where $\beta$ is a specific element of ${\mathbb F}_d$ and each $x_j$ is identified with an element of ${\mathbb Z}_r$.  We can think of $\beta$ as a root of an $n^{th}$ degree polynomial with coefficients in ${\mathbb Z}_r$
that does not factor in ${\mathbb Z}_r$.  (In a similar way, we construct the complex numbers by defining $i$ to be a root of $x^2 + 1$, which does not factor in the reals.)  
Eq.~(\ref{vectorspace}) shows that we may regard ${\mathbb F}_d$ as an $n$-dimensional vector space over ${\mathbb Z}_r$.  Of course it is much more than that, since its elements can also be multiplied. 
In this paper we will make essential use of the notion of the {\em trace} of a field element, which can be used to map a general element of ${\mathbb F}_d$ into
an element of ${\mathbb Z}_r$.  The trace of $y$, with $y \in {\mathbb F}_d$, is defined
by
\begin{equation}  \label{trace}
\hbox{tr}\,y = y + y^r + y^{r^2} + \cdots + y^{r^{n-1}}.
\end{equation}
(We use the lower-case ``tr'' to distinguish the field trace from the trace of a matrix.)  
Though it is not obvious from the definition, $\hbox{tr}\,y$ is a field element $x$ for which all the $x_j$'s in Eq.~(\ref{vectorspace}) are equal to zero except possibly $x_0$.  Eq.~(\ref{trace}) therefore identifies an element of ${\mathbb Z}_r$.  
The trace has the following properties:
\begin{equation}
\hbox{tr}(y+z) = \hbox{tr}\,y + \hbox{tr}\,z \hspace{3mm} \hbox{for $y$ and $z$ in ${\mathbb F}_d$},
\end{equation}
and, again regarding ${\mathbb F}_d$ as a vector space over ${\mathbb Z}_r$,
\begin{equation}
\hbox{tr}(ay) = a\,\hbox{tr}\,y \hspace{3mm} \hbox{for $y \in {\mathbb F}_d$ and $a \in {\mathbb Z}_r$}.
\end{equation}
That is, the trace defines a linear map from the $n$-dimensional vector space to ${\mathbb Z}_r$ itself.

We can now define our discrete Wigner function.  Let $r$ be an odd prime and let $d$ be equal to $r^n$ for some positive integer $n$.  For any complex Hermitian $d \times d$ matrix $R$, the discrete Wigner function $W_R$ associated with $R$ is a real function of 
the phase space point $(q,p)$, where again $q$ and $p$ take values in ${\mathbb F}_d$.  (We will think of 
$q$ as the horizontal coordinate and $p$ as the vertical coordinate.)  $W_R$ is defined as\cite{Klimov1, Vourdas}
\begin{equation}  \label{Wigdef}
W_R(q,p) = \frac{1}{d}\,\hbox{Tr}\left[RA(q,p)\right],
\end{equation}
where $A(q,p)$ is the $d \times d$, unit-trace, Hermitian matrix given by
\begin{equation}  \label{Adef}
[A(q,p)]_{jk} = \delta_{j,2q - k}\omega^{\hbox{\scriptsize tr}[(j-k)p]}.
\end{equation}
Here $\omega$ is the $r^{th}$ root of unity $\omega = e^{2 \pi i/r}$ and the indices $j$ and $k$ take values in ${\mathbb F}_d$.
The arithmetic in the argument of the Kronecker delta and in the exponent is in ${\mathbb F}_d$, and the trace is to be interpreted as an ordinary integer
exponent.
The usefulness of the choice (\ref{Adef}) of the matrices $A$ will become clear later in this section.  

We will frequently use the following identity, which generalizes a familiar fact about roots of unity: for any $y \in {\mathbb F}_d$, 
\begin{equation} \label{omegasum}
\sum_{x \in {\mathbb F}_d} \omega^{\hbox{\scriptsize tr}(xy)} = d\delta_{y,0}.
\end{equation}
One consequence of this identity is that the $A$'s are orthonormal in the sense that 
\begin{equation}  \label{ortho}
\hbox{Tr}\left[A(q_1,p_1)A(q_2,p_2)\right] = d\delta_{q_1,q_2}\delta_{p_1,p_2}.
\end{equation}
Note also that there are $d^2$ of these matrices; so they constitute a complete basis for the space of $d \times d$ Hermitian matrices.  
From Eq.~(\ref{ortho}) and the definition (\ref{Wigdef}) we can write $R$ in terms of $W_R$:
\begin{equation}  \label{Wiginverse}
R = \sum_{q,p} W_R(q,p) A(q,p).
\end{equation}
That is, the numbers $W_R(q,p)$ are the coefficients in the expansion of $R$ as a linear combination of the $A$'s.

Two properties of the Wigner function will be particularly useful for our purposes.  First, because of the orthonormality 
of the $A$ matrices, we have, for any Hermitian $R$ and $S$,
\begin{equation}  \label{innerproduct}
\hbox{Tr}(RS) = d \sum_{q,p} W_R(q,p)W_S(q,p).
\end{equation}
The other property is the rule for finding the Wigner function of a product $RS$, given the Wigner functions of $R$ and $S$.
From the above definitions one can work out that this rule is given as follows:
\begin{equation}  \label{productrule}
\begin{split}
W_{RS}(q_1,p_1) = \sum_{q_2,p_2,q_3,p_3} \Gamma(q_1,p_1,q_2,p_2,q_3,p_3)
 W_R(q_2,p_2)W_S(q_3,p_3),
\end{split}
\end{equation}
where 
\begin{equation}  \label{Gamma}
\Gamma(q_1,p_1,q_2,p_2,q_3,p_3) = \frac{1}{d}\,\omega^{\hbox{\scriptsize tr}\left\{2\left[(q_3-q_2)p_1 + (q_1 - q_3)p_2 + (q_2 - q_1)p_3\right]\right\} }.
\end{equation}

As we have said, in discrete phase space one can speak of lines and parallel lines: A line is the set of solutions $\{(q_1,p_1), \ldots, (q_d, p_d)\}$
of an equation of the form $a q + b p = c$, where $a$, $b$, and $c$ are elements of 
${\mathbb F}_d$ with $a$ and $b$ not both zero.  Two lines are called parallel if they can be expressed by two such linear equations differing only in the value of 
$c$.  There are in total $d(d+1)$ lines, which can be grouped into $d+1$ sets of $d$ parallel lines; 
these sets of lines are the striations of the discrete phase space.  The $d+1$ striations correspond to the $d+1$ possible
slopes of the lines, that is, the possible values of $-a/b$.  These slope values include all the elements of ${\mathbb F}_d$ along with $\infty$ (infinite slope 
corresponding to the case $b = 0$).  Note that two lines that are not parallel intersect in exactly one point.  

Now we make the connection with mutually unbiased bases.
For each line, consider the function on phase space that is nonzero only on that line, where it has the constant
value $1/d$.  Starting from Eq.~(\ref{Wiginverse}), it is not hard to show that this function is the Wigner function 
of a pure-state density matrix.  This property is in fact the main motivation 
for choosing the form (\ref{Adef}) of the $A$ matrices.  Thus every line in phase space corresponds to a pure state.  
It then follows from Eqs.~(\ref{ortho}) and (\ref{innerproduct}) that the states corresponding to parallel lines are
orthogonal.  Since there are $d$ parallel lines in a striation, each striation corresponds to an orthogonal basis
for the Hilbert space.  

Now consider two state vectors, corresponding to two lines that belong to different striations.  Let $\rho_1$ and $\rho_2$
be the density matrices of the two states, and let $\lambda_1$ and $\lambda_2$ be the corresponding lines in phase space.
Then we have
\begin{equation}
\begin{split}
\hbox{Tr}(\rho_1\rho_2) &= \hbox{Tr}\left(\frac{1}{d}\sum_{(q_1,p_1) \in \lambda_1}A(q_1,p_1)\right)\left(\frac{1}{d}\sum_{(q_2,p_2) \in \lambda_2}A(q_2,p_2)\right)\\
&= \frac{1}{d^2} \sum_{(q_1,p_1) \in \lambda_1}\sum_{(q_2,p_2) \in \lambda_2} \hbox{Tr}\left[A(q_1,p_1)A(q_2,p_2)\right].
\end{split}
\end{equation}
But the $A$'s are orthogonal, and there is exactly one point that is common to both lines, so there is only one nonzero term in the sum.  According to Eq.~(\ref{ortho}) the value of this term is $d$.  Therefore
\begin{equation}
\hbox{Tr}(\rho_1\rho_2) = \frac{1}{d}.
\end{equation}
Thus the orthogonal bases associated with two distinct striations have the property that if we choose any two vectors, one from each basis, their
inner product will always have the same magnitude, $1/\sqrt{d}$---the bases are mutually unbiased.  

Earlier we claimed that the sums of the Wigner function over the lines of a striation are the probabilities of the outcomes
of the measurement associated with that striation.  To see why this is true, let $\rho_\lambda$ be the density matrix of the pure state associated with 
the line $\lambda$, and let $\sigma$ be the density matrix of the state being measured, whose Wigner function is $W_\sigma$.  
Then
\begin{equation}
\sum_{(q,p) \in \lambda} W_\sigma(q,p) = 
\frac{1}{d}\sum_{(q,p) \in \lambda} \hbox{Tr}\left[\sigma A(q,p)\right]
=\hbox{Tr}\left[\sigma\rho_\lambda\right],
\end{equation}
which is indeed the probability of obtaining the outcome corresponding to the pure state $\rho_\lambda$.

We now specialize to the case $d=3$ (mod 4) and write down the Wigner function of the state we claim is MUB-balanced.  We call this Wigner function $W_{\rho}$, anticipating that it is indeed
the Wigner function corresponding to a legitimate density matrix $\rho$, but we will have to prove this.  We arrived at $W_\rho$ via methods
developed by Sussman\cite{Sussmanthesis} and Appleby\cite{Applebypublished}, but our proofs will not depend on how the state was derived.
We will be able to show from the form of the Wigner function itself that it satisfies the conditions of a MUB-balanced state.  
The Wigner function is
\begin{equation}  \label{specialWig}
W_\rho(q,p) = \frac{1}{d(d+1)}\left[1-d\delta_{q, 0}\delta_{p,0}+\sum_{x \in {\mathbb F}_d^*} \eta(x^2+1)\omega^{\hbox{\scriptsize tr} [x(q^2+p^2)]} \right],
\end{equation}
where ${\mathbb F}_d^*$ consists of all the nonzero elements of ${\mathbb F}_d$, and the function $\eta$ is the quadratic character: for $y \in {\mathbb F}_d^*$, $\eta(y)$ is defined by
\begin{equation}
\eta(y) = \left\{ \begin{array}{rl}+1&\hbox{if $y = s^2$ for some $s \in {\mathbb F}_d^*$} \\
-1 &\hbox{if $y \ne s^2$ for any $s \in {\mathbb F}_d^*$}
\end{array} \right.
\end{equation}
We will never encounter $\eta(0)$.  In Eq.~(\ref{specialWig}), the argument $x^2 + 1$ of $\eta$ will never equal zero because negative one has no square root in ${\mathbb F}_d$.  (See item 3 in the list below.)  

Notice that $W_\rho$ is a real function: in the sum over $x$ the terms $\omega^{\hbox{\scriptsize tr}[x(q^2+p^2)]}$ and $\omega^{\hbox{\scriptsize tr}[-x(q^2+p^2)]}$, which are conjugates of each other,
are multiplied by the same factor.  Therefore $W_\rho$ is the Wigner function of some Hermitian operator $\rho$ in accordance with
Eq.~(\ref{Wiginverse}).  
We need to show (i) that $\rho$ is the density matrix of a pure state, and (ii) that this state is MUB-balanced.  We prove these statements 
in the next two sections.  

Before we get into the proofs, it may be helpful to gather at this point a few algebraic facts that we will use in the following sections.\cite{Lidl}  These first two facts
apply to any odd $d = r^n$ with $r$ prime:
\begin{enumerate}
\item For all $x, y \in {\mathbb F}_d^*$, $\eta(xy) = \eta(x)\eta(y)$.
\item $\sum_{x \in {\mathbb F}_d^*}\eta(x^2 + 1) = -2$. (See Lidl and Niederreiter\cite{Lidl}, p.~230.)
\end{enumerate}
In addition, when $d$ is equivalent to 3 (mod 4), we have the following:
\begin{enumerate}
\setcounter{enumi}{2}
\item The field element $-1$ is not the square of any element.  That is, $\eta(-1) = -1$.
\item It follows that multiplying $x$ by $-1$ changes the sign of $\eta(x)$.  
\item For $x \in {\mathbb F}_d^*$, $\sum_{q \in {\mathbb F}_d} \omega^{\hbox{\scriptsize tr}[xq^2]}= i^n\eta(x)\sqrt{d}$. 
(See Lidl and Niederreiter\cite{Lidl}, p.~218.)  
Note that 
when $d = 3$ (mod 4), the exponent $n$ in $d = r^n$ must be odd.  So this sum is purely imaginary.

\end{enumerate}

\section{Proof that $W_\rho$ is the Wigner function of a pure state}  \label{secthree}

To show that $\rho$ is a pure-state
density matrix, it is sufficient to show that $\rho^2 = \rho$ and that $\hbox{Tr}\,\rho = 1$.  The latter condition will be true
if the sum of $W_\rho(q,p)$ over all phase-space points $(q,p)$ is equal to 1 (as follows from Eq.~(\ref{Wiginverse}) and
the fact that $A(q,p)$ has unit trace).  Let us evaluate this sum for $W_\rho$ as given in Eq.~(\ref{specialWig}):
\begin{equation} \label{Wsum}
\begin{split}
\sum_{q,p} W_\rho(q,p) &= \frac{1}{d(d+1)}
 \left[ d^2 - d + \sum\limits_{x \in {\mathbb F}_d^*}\eta(x^2+1)\Bigg(\sum\limits_{q \in {\mathbb F}_d}\omega^{\hbox{\scriptsize tr}[xq^2]}\Bigg)^{\hspace{-1mm}2}\right].
\end{split}
\end{equation}
Here we have used the fact that $\sum_{q,p}\omega^{\hbox{\scriptsize tr}[x(q^2 + p^2)]}$ is the square of a single sum.  Now, for
$d=3$ (mod 4), the sum over $q$ in Eq.~(\ref{Wsum}) is equal to $\pm i\sqrt{d}$.  Thus when we square this sum we get simply $-d$.  This leaves
the sum over $x$, that is, $\sum_x \eta(x^2 + 1)$, which is equal to $-2$.
We therefore have
\begin{equation}
\sum_{q,p} W_\rho(q,p) = \frac{1}{d(d+1)}\left[ d^2 - d + 2d \right] = 1.
\end{equation}
So the Hermitian matrix $\rho$ represented by $W_\rho$ does indeed have unit trace.  

Showing that $\rho^2 = \rho$ requires more work.  We prove it by proving the equivalent statement for the Wigner function, based on
Eq.~(\ref{productrule}). That is, we will show that
\begin{equation}  \label{nastysum}
W_\rho(q_1,p_1) = \sum_{q_2,p_2,q_3,p_3} \Gamma(q_1,p_1,q_2,p_2,q_3,p_3) W_\rho(q_2,p_2)W_\rho(q_3,p_3).
\end{equation}

Let $S(q_1,p_1)$ be the sum on the right-hand side of Eq.~(\ref{nastysum}).  We want to show that 
$S(q_1,p_1) = W_\rho(q_1,p_1)$.  We will do the sum by breaking it into parts.  The Wigner function of Eq.~(\ref{specialWig}) has three terms
inside the square bracket; let us call them $T_1$, $T_2$, and $T_3$---that is, $T_1 = 1$, $T_2 = -d\delta_{q,0}\delta_{p,0}$, and $T_3$ is the sum over $x$.  We define the following functions $X_{mn}$ that arise from these terms when we do the operations in Eq.~(\ref{nastysum}):
\begin{equation}
X_{mn}(q_1,p_1) = \sum_{q_2,p_2,q_3,p_3}\omega^{\hbox{\scriptsize tr}\left\{2\left[(q_3-q_2)p_1 + (q_1 - q_3)p_2 + (q_2 - q_1)p_3\right]\right\}} T_m(q_2,p_2)T_n(q_3,p_3).
\end{equation}
In terms of the $X$'s, the desired sum is
\begin{equation}  \label{S}
S = \frac{1}{d^3(d+1)^2}\left[ X_{11} + X_{22} + X_{33} + 2\,\hbox{Re}(X_{12} + X_{13} + X_{23})\right].
\end{equation}
For the cross terms, we get twice the real part because interchanging $(q_2,p_2)$ with $(q_3,p_3)$ has the 
effect of complex conjugating $\Gamma$.  (It will turn out, though, that each $X_{mn}$ is already real.)
For four of the terms in Eq.~(\ref{S}) the 
evaluation is straightforward and we simply present the results here:
\begin{equation}  \label{X11etc}
\begin{split}
X_{11}(q,p) &= d^2. \\
X_{12}(q,p) &=  -d^3 \delta_{q,0}\delta_{p,0}.  \hfill \\
X_{13}(q,p) &=  d^2  \sum_{x \in {\mathbb F}_d^*}\eta(x^2+1) \omega^{\hbox{\scriptsize tr}[x(q^2 + p^2)]}.  \\
X_{22}(q,p) &=  d^2.
\end{split}
\end{equation}
We now go through the details for the other two terms, $X_{23}$ and $X_{33}$.

\bigskip

\noindent $\underline{X_{23}}$:

\bigskip

\noindent Here the two delta functions have the effect of setting $q_2$ and $p_2$ equal to zero; so the remaining sum is
\begin{equation}
X_{23}(q_1,p_1) = -d \sum_{x \in {\mathbb F}_d^*}\eta(x^2+1) \sum_{q_3,p_3} \omega^{\hbox{\scriptsize tr}[2(q_3p_1  - q_1p_3)]}\omega^{\hbox{\scriptsize tr}[x(q_3^2 + p_3^2)]}.
\end{equation}
By completing the squares and shifting the summation variables $q_3$ and $p_3$, one can write this as
\begin{equation}  \label{X23}
\begin{split}
X_{23}(q_1,p_1) &= -d \sum_{x \in {\mathbb F}_d^*}\eta(x^2+1) \sum_{q_3,p_3} \omega^{\hbox{\scriptsize tr}[x(q_3^2 + p_3^2)]}\omega^{\hbox{\scriptsize tr}[-\frac{1}{x}(q_1^2 + p_1^2)]} \\
&= d^2  \sum_{x \in {\mathbb F}_d^*}\eta(x^2+1)\omega^{\hbox{\scriptsize tr}[-\frac{1}{x}(q_1^2 + p_1^2)]} \\
&=d^2  \sum_{x \in {\mathbb F}_d^*}\eta(x^2+1)\omega^{\hbox{\scriptsize tr}[x(q_1^2 + p_1^2)]}.
\end{split}
\end{equation}
The last step can be justified by changing the summation variable to $y = -1/x$.  Then the argument of $\eta$ becomes $(-1/y)^2 + 1$.  But $(-1/y)^2 + 1 = 
(y^2 + 1)(1/y^2)$.  So $\eta((-1/y)^2 + 1) = \eta(y^2 + 1)\eta(1/y^2) = \eta(y^2 + 1)$.  

\bigskip

\noindent $\underline{X_{33}}$:
\begin{equation}  \label{X33}
X_{33}(q_1,p_1) =  \sum_{x \in {\mathbb F}_d^*} \sum_{y \in {\mathbb F}_d^*}\eta(x^2+1) \eta(y^2+1) f(x,y,q_1,p_1),
\end{equation}
where
\begin{equation}
\begin{split}
f(x,y,&q_1,p_1) =  \sum_{q_2,p_2,q_3,p_3}\omega^{\hbox{\scriptsize tr}\left\{2\left[(q_3-q_2)p_1 + (q_1 - q_3)p_2 + (q_2 - q_1)p_3\right]\right\}}\omega^{\hbox{\scriptsize tr}[x(q_2^2 + p_2^2)]}\omega^{\hbox{\scriptsize tr}[y(q_3^2 + p_3^2)]}\\
= &\sum_{q_2,p_3}\omega^{\hbox{\scriptsize tr}[2(-q_2p_1+q_2p_3 - q_1p_3) +xq_2^2 + yp_3^2]}
\sum_{q_3,p_2}\omega^{\hbox{\scriptsize tr}[2(q_3p_1+q_1p_2-q_3p_2) + xp_2^2 + yq_3^2]}.
\end{split}
\end{equation}
These sums can be done by completing squares.  One has to distinguish two cases: (i) $y = 1/x$, and (ii) $y \ne 1/x$.  In the first case, the result comes out to be
\begin{equation}
f(x,y,q_1,p_1) = -d^3 \delta_{q_1,0}\delta_{p_1,0}.     \hspace{1in}  \left( y = \frac{1}{x} \right)
\end{equation}
And in the second case, one gets
\begin{equation}
f(x,y,q_1,p_1) = d^2 \omega^{\hbox{\scriptsize tr}\{\left[(x+y)/(1-xy)\right](q_1^2 + p_1^2)\}}. \hspace{0.4in}  \left( y \ne \frac{1}{x} \right)
\end{equation}
We now plug these expressions back into Eq.~(\ref{X33}).  Let us write $X_{33} = X_{33}^{(1)} + X_{33}^{(2)}$, where
the first part includes all the terms with $y = 1/x$, and the second includes all the rest.  In the sum over $x$ and $y$, there
are $d-1$ terms with $y=1/x$, and they all have the same value.  So we have
\begin{equation}
X_{33}^{(1)} = - d^3(d-1)\delta_{q_1,0}\delta_{p_1,0}.
\end{equation}
The remaining part is
\begin{equation}
X_{33}^{(2)}=d^2 \sum_{x \in {\mathbb F}_d^*} \sum_{\begin{array}{c}\vspace{-10.5mm}\\{\hbox{\scriptsize $y\in {\mathbb F}_d^*$}}\\ \vspace{-10.6mm}\\{\hbox{\scriptsize $y\ne1/x$}}\end{array}}\hspace{-2mm}\eta(x^2+1)\eta(y^2+1)
\omega^{\hbox{\scriptsize tr}\{\left[(x+y)/(1-xy)\right](q_1^2 + p_1^2)\}} .
\end{equation}
For brevity, we now use the symbol $z$ for the combination $(x+y)/(1-xy)$.  Fortunately, the value of $\eta(z^2+1)$ is
the same as the value of $\eta[(x^2+1)(y^2+1)]$:
\begin{equation}
z^2 +1 = \left[\frac{x+y}{1-xy}\right]^2 + 1 = \frac{(x^2+1)(y^2+1)}{(1-xy)^2},
\end{equation}
so that the ratio is a perfect square.  We can therefore write
\begin{equation}  \label{X332}
X_{33}^{(2)}=d^2 \sum_{x \in {\mathbb F}_d^*} \sum_{\begin{array}{c}\vspace{-10.5mm}\\{\hbox{\scriptsize $y\in {\mathbb F}_d^*$}}\\ \vspace{-10.6mm}\\{\hbox{\scriptsize $y\ne1/x$}}\end{array}}\hspace{-2mm}\eta(z^2+1)
\omega^{\hbox{\scriptsize tr}[z(q_1^2 + p_1^2)]} .
\end{equation}

Now, for a given value of $z$, how many allowed pairs $(x,y)$ yield that value of $z$?  For the special case $z=0$, there are $d-1$ such pairs, namely, all
those for which $x = -y$.  For any other value of $z$, there are $d-3$ such pairs.  To see this, solve for $y$ in terms of $z$ and $x$:
\begin{equation}
y = \frac{z-x}{1+zx}.
\end{equation}
If $x\ne z$ and $x\ne -1/z$, then there is exactly one allowed value of $y$ that gives the desired $z$.  If $x=z$ or $x=-1/z$, there is
no such value.  So the number of terms is $(d-1) - 2 = d-3$.  

We can therefore rewrite Eq.~(\ref{X332}) as
\begin{equation}
X_{33}^{(2)} = d^2(d-1) + d^2(d-3)\sum_{z \in {\mathbb F}_d^*}\eta(q^2+1)\omega^{\hbox{\scriptsize tr}[z(q_1^2 + p_1^2)]}.
\end{equation}
The whole term $X_{33}$ can thus be written as
\begin{equation}  \label{X33final}
X_{33} =  - d^3(d-1)\delta_{q_1,0}\delta_{p_1,0}+d^2(d-1) + d^2(d-3)\sum_{z \in {\mathbb F}_d^*}\eta(z^2+1)\omega^{\hbox{\scriptsize tr}[z(q_1^2 + p_1^2)]}.
\end{equation}

\bigskip

\noindent{\em Putting the pieces together}

\bigskip

We can now evaluate the right-hand side of Eq.~(\ref{S}) by collecting the results expressed in Eqs.~(\ref{X11etc}, \ref{X23}, \ref{X33final}).  We get
\begin{equation}
S(q,p) = \frac{1}{d(d+1)}\left[1-d\delta_{q, 0}\delta_{p, 0}+\sum\limits_{x \in {\mathbb F}_d^*}\eta(x^2+1)\omega^{\hbox{\scriptsize tr}[x(q^2+p^2)]}\right],
\end{equation}
which is indeed the original Wigner function of Eq.~(\ref{specialWig}).  So we have shown that $\rho^2 = \rho$.  According to what we
have said before, it follows that $\rho$ is the density matrix of a pure state.

\section{Proof that the state $\rho$ is MUB-balanced}  \label{secfour}

We now want to show that when the Wigner function $W_\rho$ of Eq.~(\ref{specialWig}) is summed over the lines of any striation, we always get the
same list of values up to permutations.  This will show that the state represented by $W_\rho$ is MUB-balanced.  For this proof we need only two facts
about $W$: (i) $W_\rho$ is of the form $W_\rho(q,p) = f(q^2 + p^2)$, and (ii) the function $f(s)$ has the property that $f(-s) = f(s)$.   To see
that the latter property holds, note that in Eq.~(\ref{specialWig}), we can cancel a factor of $-1$ in the exponent of $\omega$ by changing the 
summation variable to $y = -x$.  

Let us first consider a striation with a slope $m$ that is not infinity.  Let the lines of the striation be defined by the linear equations
\begin{equation}
p = mq + b,
\end{equation}
where the ``vertical displacement'' $b$ can take any value in ${\mathbb F}_d$.  The $d$ lines of the striation are distinguished from each other only by the value of $b$.  Let $\lambda_b$ be the line in this striation with vertical displacement $b$.  
We write our special Wigner function simply as
\begin{equation}
W_\rho(q,p) = f(q^2 + p^2).
\end{equation}
When we sum this function over $\lambda_b$, we get
\begin{equation}
\sum_{(q,p) \in \lambda_b} W_\rho(q,p) = 
\sum_{q \in {\mathbb F}_d} f\hspace{-1mm}\left[q^2 + (mq + b)^2\right]
= \sum_{q \in {\mathbb F}_d} f\hspace{-1mm}\left[(m^2+1)q^2 + 2mbq + b^2 \right].
\end{equation}
By completing the square and shifting the summation variable $q$, we can bring this expression to the form
\begin{equation} \label{AAA}
\sum_{(q,p) \in \lambda_b} W_\rho(q,p) =
\sum_{q \in {\mathbb F}_d} f\hspace{-1mm}\left[(m^2+1)q^2 + \frac{b^2}{m^2+1}\right].
\end{equation}
If $\eta(m^2+1)$ is equal to 1---that is, if $m^2 +1 = t^2$ for some nonzero $t$---then we can define a new summation variable $q' = tq$, so that
\begin{equation}  \label{qprime}
\sum_{(q,p) \in \lambda_b} W(q,p) =
\sum_{q' \in {\mathbb F}_d} f\hspace{-1mm}\left[q'^2 + \frac{b^2}{m^2+1}\right].
\end{equation}
Now as $b$ ranges over
the values in ${\mathbb F}_d$, the term $b^2/(m^2+1)$ takes the value zero once, and it takes each nonzero value that is a perfect square exactly twice.
Thus every value of $m$ for which $\eta(m^2 + 1) = 1$ yields the same list of probabilities (up to permutations).  
On the other hand, if $\eta(m^2 + 1) = -1$, we can use the symmetry of the function $f$ to rewrite Eq.~(\ref{AAA}) as
\begin{equation}
\sum_{(q,p) \in \lambda_b} W(q,p) =
\sum_{q \in {\mathbb F}_d} f\hspace{-1mm}\left[\left[-(m^2+1)\right]q^2 + \frac{b^2}{[-(m^2+1)]}\right].
\end{equation}
But now $\eta[-(m^2+1)]$ {\em is} equal to 1, and so we get the same set of values as before. 
Thus for every value of $m$ other than infinity, we get the same set of probabilities of the outcomes of the corresponding measurement.  

It is not hard to check that $m=\infty$ also yields the same set of values.  In that case, let the lines of the striation be defined by the 
equations $q = b$ with $b \in {\mathbb F}_d$.  Summing over a line simply means summing over $p$.  Then
\begin{equation}
\sum_{(q,p) \in \lambda_b} W(q,p) = 
\sum_{p \in {\mathbb F}_d} f(p^2 + b^2),
\end{equation}
which is the same as Eq.~(\ref{AAA}) with $m=0$.  So $m=\infty$ yields the same probability values.  

We thus see that summing our special Wigner function $W_\rho$ over the lines of any striation yields the same list of probabilities up to 
permutation.  So $W_\rho$ represents a MUB-balanced state.  

Note that one of these shared probabilities is zero: for each of the striations, the probability associated with the line through the origin
is (see Eqs.~(\ref{specialWig}) and (\ref{qprime}))
\begin{equation}
\begin{split}
\sum_{(q,p) \in \lambda_0} W_\rho(q,p) = \sum_{q \in {\mathbb F}_d} f\hspace{-1mm}\left(q^2\right) &= \frac{1}{d(d+1)}\sum_{q \in {\mathbb F}_d}
\left[1-d\delta_{q, 0}+\sum\limits_{x \in {\mathbb F}_d^*}\eta(x^2+1)\omega^{\hbox{\scriptsize tr}(xq^2)}\right] \\
&= \frac{1}{d(d+1)}\left(i^n\sqrt{d}\right)\sum_{x \in {\mathbb F}_d^*} \eta(x)\eta(x^2 + 1),
\end{split}
\end{equation}
which is zero because the summand is odd in $x$.  

It is worth pointing out an interesting difference between the continuous phase space and our discrete phase space with $d=3$ (mod 4).
In the former case we can define a circle to be the set of solutions to an equation of the form $q^2 + p^2 = c$, where $c$ is some nonzero real constant.
Of course $c$ cannot be just {\em any} nonzero real constant if the equation is to have a solution: it must be positive.  In our discrete phase space, we can again define a circle to be the set of solutions
to an equation of the form $q^2 + p^2 = c$ (with arithmetic in ${\mathbb F}_d$), but now any nonzero value of $c$ allows a solution, and there are two kinds of circle: those for which $\eta(c)=1$,
and those for which $\eta(c) = -1$.  
(Note that as an alternative to the polynomial $q^2 + p^2$ in our definition of ``circle," we could, with just as much justification, use
any other non-factorable homogeneous polynomial of degree 2 in $q$ and $p$.  There is no natural notion of ``distance'' in ${\mathbb F}_d^2$.  However, the polynomial $q^2 + p^2$ is convenient because
when $d$ is equal to 3 (mod 4) it is guaranteed to be non-factorable.)
To prove that a state is MUB-balanced, it is not enough to know that its
Wigner function is constant on every circle (as, in the continuous case, an energy eigenstate of a harmonic oscillator is constant on
every circle).  That is, it is not enough that $W_\rho$ depends on $q$ and $p$ only through the combination
$q^2 + p^2$.  In the above argument we also needed the fact that $W_\rho$ takes the same value on the circle $q^2 + p^2 = -c$ as it does
on the circle $q^2 + p^2 = c$.  That is, for each circle of one kind, there needed to be a circle of the other kind with a matching value of 
the Wigner function.  

Now that we have identified one MUB-balanced state for each of our values of $d$, it is not hard to generate others.  Let $L$ be a unit-determinant
$2 \times 2$ matrix with entries in ${\mathbb F}_d$.  Then $L$ takes each phase space point $(q, p)$ into a phase space point $(q', p')$
according to 
\begin{equation}
L \mtx{c}{q \\ p} = \mtx{c}{q' \\ p'}.
\end{equation}
Now let the Wigner function $W_{\rho'}$ be defined by
\begin{equation}
W_{\rho'}(q,p) = W_\rho(q', p').
\end{equation}
Then $W_{\rho'}$ also represents a MUB-balanced state, as we now show.  First, that $W_{\rho'}$ represents a pure state is guaranteed by a correspondence between unit-determinant linear transformations on 
phase space (for odd prime-power $d$) and certain unitary operators.\cite{linear, Gross1, Applebypublished}  In effect, we are simply performing a unitary transformation on our original MUB-balanced state; so the result is certainly a pure state.
Second, the linear transformation preserves lines in phase space and preserves the notion of parallel lines.  So the new Wigner function $W_{\rho'}$ can be pictured as a permutation in phase space of the values
of the original Wigner function, but it is a permutation that respects the striation structure.  Thus the list of probabilities arising from summing $W_{\rho'}$ over any set of parallel lines matches those arising from summing $W_\rho$ over
a (possibly different) set of parallel lines.  It follows that $W_{\rho'}$ is MUB-balanced if $W_\rho$ is MUB-balanced.  This use of linear transformations is analogous to a squeezing operation on the continuous phase space.  

In a similar way, we can generate yet more MUB-balanced states through {\em translations} of the phase space, which are likewise associated with unitary transformations on the Hilbert space.\cite{Applebypublished, Klimov1, Vourdas}
(The continuous analog would be a displacement in the continuous phase space.)  Starting from a single MUB-balanced state, the collection of states generated by applying to that state all possible phase-space translations defines a non-orthogonal measurement (that is, a positive-operator-valued measure), each of whose
outcomes can be identified with a MUB-balanced state.  This measurement thus bears essentially the same relation to each of the mutually unbiased bases.  

Note that all of the transformations we have mentioned here leave the set of probability values unchanged.  
We have found no set of MUB-balanced states that would be analogous to a set of distinct energy eigenstates of
a harmonic oscillator, whose probability distributions would also be quite distinct.  It is conceivable that
the set of probabilities associated with the special state defined in Eq.~(\ref{specialWig}) is the {\em only} set of probability values that can arise from a MUB-balanced state in dimension $d$.

\section{The density matrix and the state vector}  \label{secfive}

Eq.~(\ref{Wiginverse}) tells us how to construct the density matrix corresponding to a given Wigner function.  For our special Wigner
function specified in Eq.~({\ref{specialWig}), this formula gives us the components of $\rho$, that is, $\rho_{jk} \equiv \langle b^{(0)}_j|\rho|b^{(0)}_k\rangle$, where $\{|b^{(0)}_j\rangle | j \in {\mathbb F}_d\}$ is the
standard basis.
\begin{equation}
\rho_{jk} = \sum_{q,p \in {\mathbb F}_d}  \frac{1}{d(d+1)}\left[1-d\delta_{q, 0}\delta_{p,0}+\sum\limits_{x \in {\mathbb F}_d^*}\eta(x^2+1)\omega^{\hbox{\scriptsize tr}[x(q^2+p^2)]}\right] 
 \left( \delta_{j,2q - k}\omega^{\hbox{\scriptsize tr}[(j-k)p]}\right).
\end{equation}
The sum over $q$ is straightforward, and the sum over $p$ can be done by completing the square in the exponent.  The result is
\begin{equation}
\rho_{jk} = \frac{1}{d+1}\left[ \delta_{j,k} - \delta_{j,-k} + \frac{i^n}{\sqrt{d}} \sum_{x \in {\mathbb F}_d^*}\eta(x)\eta(x^2+1)
\omega^{\hbox{\scriptsize tr}\left\{(1/4)\left[x(j+k)^2 - (1/x)(j-k)^2\right]\right\}} \right].
\end{equation}
This density matrix is entirely real: the factor $i^n$ is imaginary, and the terms in the sum corresponding to $x$ and $-x$ are the negative complex conjugates of each other, since $\eta(-x)
= - \eta(x)$.  We know from Section \ref{secthree} that $\rho$ is of the form $\rho = |\psi\rangle\langle\psi |$ for some normalized state vector $|\psi\rangle$.  This fact has also been proved directly
by Evans through an argument reproduced in Katz's paper\cite{Katz} (an argument that does not involve the Wigner function).
We now see that $|\psi\rangle$ can be taken to have only real components in the standard basis.  

We can obtain the vector $|\psi\rangle$ from the above expression for $\rho$.  For any fixed value of $k$, we can say
that $\psi_j \equiv \langle b^{(0)}_j|\psi\rangle  \propto \langle b^{(0)}_j|\psi\rangle\langle \psi | b^{(0)}_k\rangle=\langle b^{(0)}_j|\rho| b^{(0)}_k\rangle = \rho_{jk}$, and as long as this last vector (with $k$ fixed) is not the zero vector, we can obtain $|\psi\rangle$ by normalizing it.  (We cannot use the value
$k=0$ in this way, because $\rho_{j0}$ is indeed the zero vector.  This follows from the fact---seen in the preceding section---that the sum 
of $W_\rho(q,p)$ over the line $q=0$ is equal to
zero.  The measurement outcome associated with the vertical line $q=0$ is the one whose probability is $\psi_0^2$.)  

It turns out to be interesting to look at a histogram of the values of the components $\psi_j$.  We show an example in Fig.~1; the dimension
in that example is $d=22307$ (the 2500$^{th}$ prime) and we have plotted a histogram of the components of the larger vector $\sqrt{d}|\psi\rangle$.  The distribution of the 
values of the components appears approximately semicircular.  We see a similar shape for prime-power values of $d$ (but not when $d$ is a power of 3).  As we mentioned in the Introduction, it has been proved
by Katz that the limiting distribution is indeed semicircular\cite{Katz}, as long as $d$ avoids the values $3^n$.  (A key element in his proof is the construction of an explicit expression for $|\psi\rangle$ not obtained
simply by normalizing a column of $\rho$.)  From the semicircularity we can estimate the maximum magnitude of a component of $|\psi\rangle$.  Treating the histogram as
if it were a continuous distribution $w(x)$, with $w(x)dx$ the number of components of $|\psi\rangle$ having values in the interval between $x$ and $x+dx$, let us
take the distribution to have a semicircular form:
\begin{equation}
w(x) = \alpha\sqrt{\beta^2 - x^2}, \hspace{1cm} -\beta \le x \le \beta.
\end{equation}
The values of $\alpha$ and $\beta$ can be determined by insisting (i) that the total number of components is $d$, and (ii) that the sum of the squares
of the components is 1.  That is, we insist that
\begin{equation}
\int_{-\beta}^{\beta} w(x) dx = d  \hspace{1cm} \hbox{and} \hspace{1cm} 
\int_{-\beta}^{\beta} w(x) x^2 dx = 1.
\end{equation}
From these conditions we find that $\alpha = d^2/2\pi$ and $\beta = 2/\sqrt{d}$.  Thus we expect the largest magnitude of a component to be
around $2/\sqrt{d}$.  So in a histogram of the components of the rescaled vector $\sqrt{d}|\psi\rangle$, we expect the range of values to extend roughly from $-2$ to $2$,
as indeed seems to be the case in Fig.~1.  The proof by Katz shows, in fact, that the values of the components of this rescaled vector are confined to the interval $[-2,2]$.

\begin{figure} \label{oneone}
\includegraphics[scale=0.8]{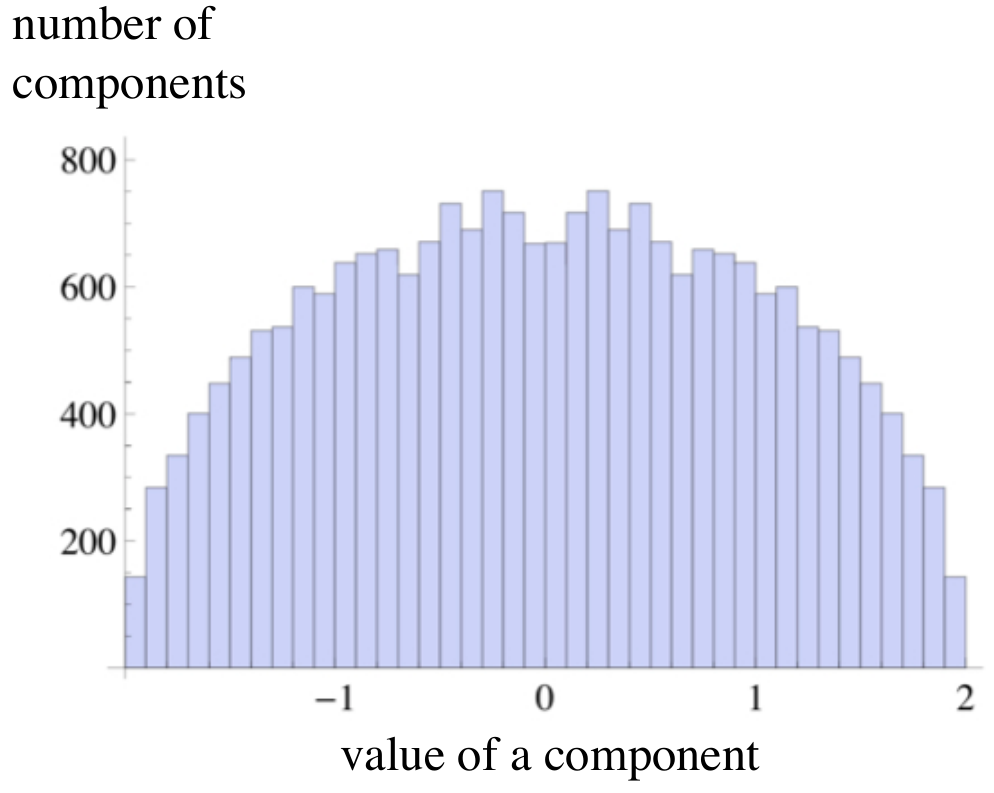}
\caption{A histogram of the values of the components of the vector $\sqrt{d}|\psi\rangle$ for $d = 22307$.  The rescaling of the vector is simply
to make the range of component values simpler.  (See the text.)}
\end{figure}


\section{Conclusions}  \label{secsix}

In this paper we have defined the notion of a state that is ``balanced'' with respect to a complete set of mutually unbiased bases.  For the complete
set of MUBs constructed from the $A$ operators of Eq.~(\ref{Adef}), we have identified, for each prime-power $d$ equivalent to 3 (mod 4), one special state that is both MUB-balanced and ``circularly symmetric''
if circles are defined by $q^2 + p^2 = c$, and we have indicated how this state can be used to generate other MUB-balanced states.  For the purpose of both specifying the state and proving that it has the desired properties, we found it easiest to work
directly with the state's discrete Wigner function, rather than with its density matrix or state vector.  From this Wigner function we obtained
an expression for the density matrix and used it to plot a histogram of the component values of the state vector, which typically
approximates a semicircle.  

As we mentioned in the Introduction, a number of previous papers have addressed the existence of minimum uncertainty states\cite{WoottersSussman, Sussmanthesis, Fuchs, Applebypublished, Galvao}, which are likewise defined relative to a 
complete set of MUBs and which are closely related to our MUB-balanced states.  
Given a complete set of MUBs $\{ |b^{(\mu)}_j\rangle \}$, {\em every} pure state $|\phi\rangle$ satisfies the inequality\cite{uncertaintyinequality}
\begin{equation}  \label{uncertainty}
\frac{1}{d+1}\sum_\mu H_2(p^{(\mu)}_1, \ldots, p^{(\mu)}_d) \ge -\log_2\left(\frac{2}{d+1}\right),
\end{equation}
where $p^{(\mu)}_j = |\langle \phi | b^{(\mu)}_j \rangle |^2$ and $H_2$ is the R\'enyi entropy of order 2:
\begin{equation}
H_2(p_1, \ldots, p_d) = - \log_2\Big( \sum_j p_j^2 \Big).
\end{equation}
The inequality follows from the convexity of the negative logarithm and the fact that\cite{two1,two2}
\begin{equation}
\sum_{\mu, j} \left(p^{(\mu)}_j\right)^2 = 2.
\end{equation}
A state is called a minimum uncertainty state if equality holds in Eq.~(\ref{uncertainty}).  This will happen whenever
the sum $\sum_j (p^{(\mu)}_j)^2$ is independent of $\mu$.  Evidently, then, any MUB-balanced state is automatically
a minimum uncertainty state, since not only the sum but the whole set of probabilities is independent of $\mu$.  
In unpublished notes, Appleby has proven the existence of at least one minimum uncertainty state in every odd prime power dimension $d$.\cite{Applebyunpublished}
However, for the case $d=1$ (mod 4), he has also shown that his construction does not yield a MUB-balanced state.  It seems to be unknown whether such states exist in these dimensions.  
 
As we suggested in our opening paragraph, there is a sense in which every energy eigenstate of a harmonic oscillator is like a MUB-balanced state.  Its probability distribution
for the observable $(\cos\theta)\hat{q} + (\sin\theta)\hat{p}$ is independent of $\theta$.  However, in that case the ``balanced'' property follows directly from the circular symmetry
of the Wigner function.  (The observables $(\cos\theta)\hat{q} + (\sin\theta)\hat{p}$, like the measurements defined by our MUBs, can be associated with striations of the phase space.\cite{WoottersWigner})  We have seen in Section \ref{secfour} that
for the discrete case, the analog of circular symmetry is not a sufficient condition to guarantee MUB-balancedness.  We also needed a symmetry between pairs of ``circles'' of the form
$q^2 + p^2 = c$ and $q^2 + p^2 = -c$.  This fact may partly explain why, for a given dimension $d$, we were able to identify only a single circularly symmetric MUB-balanced state, rather than a set of $d$ MUB-balanced states
analogous to all the energy
eigenstates of a harmonic oscillator.  The property of being MUB-balanced in finite dimensions appears to be more stringent than the analogous property in the continuous case.

\section*{Acknowledgements}

We are grateful for discussions and email correspondence
with Marcus Appleby.  We also thank Steven Miller, Ron Evans, and Nick Katz for their 
interest in the semicircular distribution suggested by our numerical results and for pursuing an explanation.  
Research by WKW is supported in part by the Foundational Questions Institute (grant FQXi-RFP3-1350).


\begin{thebibliography}{99}

\bibitem{Wigner} E.~P.~Wigner, Phys. Rev. {\bf 40}, 749 (1932).

\bibitem{Wignerreview} M.~Hillary, R.~F.~O'Connell, M.~O.~Scully, and
E.~P.~Wigner, Phys. Rep. {\bf 106}, 123 (1984).

\bibitem{WoottersWigner} W. K. Wootters, {Annals of Physics}
{\bf 176}, 1 (1987).


\bibitem{Schwinger} J. Schwinger, { Proc. Nat. Acad. Sci.}
{\bf 46}, 570 (1960).

\bibitem{Ivanovic} I. D. Ivanovic, {J. Phys. A} {\bf 14},
3241 (1981).

\bibitem{WoottersFields} W. K. Wootters and B. D. Fields, {Annals
of Physics} {\bf 191}, 363 (1989).

\bibitem{Bandyopadhyay} S. Bandyopadhyay, P. O. Boykin, 
V. Roychowdhury, and F. Vatan, Algorithmica
{\bf 34}, 512 (2002).

\bibitem{MUBreview} T.~Durt, B.-G.~Englert, I.~Bengtsson, and K.~Zyczkowski, Int.~J.~Quantum Information {\bf 8}, 535 (2010).


\bibitem{sixevidence7} M.~Grassl, arxiv:quant-ph/0406175.


\bibitem{sixevidence6} P.~Butterly and W.~Hall, Phys.~Lett.~A {\bf 369}, 5 (2007).

\bibitem{sixevidence5} S.~Brierley and S.~Weigert, Phys.~Rev.~A {\bf 78}, 042312 (2008).

\bibitem{Skinner} A.~J.~Skinner, V.~A.~Newell, and R.~Sanchez, J.~Math.~Phys.~{\bf 50}, 012107 (2009).

\bibitem{sixevidence4} P.~Raynal, X.~L\"u, B.-G.~Englert, Phys.~Rev.~A {\bf 83}, 062303 (2011).

\bibitem{sixevidence3} D.~McNulty and S.~Weigert, Int.~J.~Quant.~Inf.~{\bf 10}, 1250056 (2012).

\bibitem{sixevidence2} D.~Goyeneche, J.~Phys.~A: Math.~Theor.~{\bf 46}, 105301 (2013).

\bibitem{sixevidence1} R.~Beneduci, T.~Bullock, P.~Busch, C.~Carmeli, T. Heinosaari, and A.~Toigo,
Phys.~Rev.~A {\bf 88}, 032312 (2013).


\bibitem{QKD9} D.~Bruss, Phys.~Rev.~Lett.~{\bf 81}, 3018 (1998).


\bibitem{QKD8} H.~Bechmann-Pasquinucci and A.~Peres, Phys.~Rev.~Lett.~{\bf 85}, 3313 (2000).

\bibitem{QKD7} N.~Cerf, M.~Bourennane, A.~Karlsson, and N.~Gisin, Phys.~Rev.~Lett.~{\bf 88}, 127902 (2002).

\bibitem{QKD6} H.~F.~Chau, IEEE Trans.~Inf.~Theory {\bf 51}, 1451 (2005).

\bibitem{QKD5} I-C.~Yu, F.-L.~Lin, and C.-Y.~Huang, Phys.~Rev.~A {\bf 78}, 012344 (2008).

\bibitem{QKD4} A.~Eusebi and S.~Mancini, Quant.~Inf.~Comp.~{\bf 9}, 950 (2009).

\bibitem{QKD3} S.~Brierley, arxiv:0910.2578.


\bibitem{QKD2} M.~Mafu, A.~Dudley, S.~Goyal, D.~Giovannini, M.~McLaren, M.~J.~Padgett, T.~Konrad, F.~Petruccione, N.~L\"utkenhaus, and A.~Forbes, Phys.~Rev.~A {\bf 88}, 032305 (2013).

\bibitem{BB84} C.~Bennett, F.~Bessette, G.~Brassard, L.~Salvail, and J.~Smolin, J.~Cryptology {\bf 5}, 3 (1992).

\bibitem{Gow} R.~Gow, arxiv:math/0703333.

\bibitem{WoottersSussman}  W.~K.~Wootters and D.~M.~Sussman, in {\em Proceedings of the Eighth International 
Conference on Quantum Communication, Measurement and Computing} (NICT Press, 2007); arxiv:0704.1277.

\bibitem{Kern} O.~Kern, K.~S.~Ranade, and U.~Seyfarth, J.~Phys.~A: Math.~Theor.~{\bf 43}, 275305 (2010).

\bibitem{Seyfarth} U.~Seyfarth and K.~S.~Ranade, J.~Math.~Phys.~{\bf 53}, 062201 (2012).

\bibitem{Sussmanthesis} D.~M.~Sussman, ``Minimum-Uncertainty States and Rotational Invariance in Discrete Phase Space," Thesis, Williams College (2007).

\bibitem{Fuchs} D.~M.~Appleby, H.~B.~Dang, and C.~A.~Fuchs, arxiv:0707.2071 [quant-ph].

\bibitem{Applebypublished} D.~M.~Appleby, arxiv:0909.5233 [quant-ph].

\bibitem{Galvao} A.~Casaccino, E.~F.~Galvao, and S.~Severini, Phys.~Rev.~A {\bf 78}, 022310 (2008).



\bibitem{Applebyunpublished} D.~M.~Appleby, unpublished notes.

\bibitem{ABD} D.~M.~Appleby, I.~Bengtsson, and H.~B.~Dang, arxiv:1409.7987 [quant-ph].


\bibitem{Klimov2} A.~B.~Klimov, C.~Mu\~noz, and L.~L.~S\'anchez-Soto, Phys.~Rev.~A {\bf 80}, 043836 (2009).

\bibitem{Katz} N.~M.~Katz, Communications in Number Theory and Physics {\bf 6}, 223 (2012).  

\bibitem{BSTW} P.~O.~Boykin, M.~Sitharam, P.~H.~Tiep, and P.~Wocjan, Quantum Inf.~Comp.~{\bf 7}, 371 (2007).

\bibitem{BWB} S.~Brierley, S.~Weigert, and I.~Bengtsson, Quantum Inf.~Comp.~{\bf 10}, 803 (2010).

\bibitem{Kantor} W.~K.~Kantor, J.~Math.~Phys.~{\bf 53}, 032204 (2012).

\bibitem{discreteWigreview} C.~Ferrie, Rep.~Prog.~Phys.~{\bf 74}, 116001 (2011).

\bibitem{Klimov1} A.~B.~Klimov and C.~Mu\~noz, J.~Opt.~B: Quantum Semiclass.~Opt.~{\bf 7}, S588 (2005).

\bibitem{Vourdas} A.~Vourdas, J.~Phys.~A: Math.~Gen.~{\bf 38}, 8453 (2005).

\bibitem{Gibbons} K.~S.~Gibbons, M.~J.~Hoffman, and W.~K.~Wootters, Phys.~Rev.~A {\bf 70}, 062101 (2004)

\bibitem{Gross1} D.~Gross, J.~Math.~Phys.~{\bf 47}, 122107 (2006).

\bibitem{Gross2} D.~Gross, Appl.~Phys.~B {\bf 86}, 367 (2007).

\bibitem{Emerson1} V.~Veitch, C.~Ferrie, D.~Gross, and J.~Emerson, New J.~Phys.~{\bf 14}, 113011 (2012).

\bibitem{Emerson2} V.~Veitch, S.~A.~Hamed Mousavian, D.~Gottesman, and J.~Emerson, New J.~Phys.~{\bf 16}, 013009 (2014).


\bibitem{Lidl} R.~Lidl and H.~Niederreiter, {\em Finite Fields}, 2nd edition (Cambridge Univ.~Press, 1997).







\bibitem{linear} M.~Neuhauser, Journal of Lie Theory {\bf 12}, 15 (2002).





\bibitem{uncertaintyinequality} M.~A.~Ballester and S.~Wehner, Phys.~Rev.~A {\bf 75}, 022319 (2007).

\bibitem{two1} U.~Larsen, J.~Phys.~A {\bf 23}, 1041 (1990).

\bibitem{two2} A.~Klappenecker and M.~R\"otteler, Proceedings of the 2005 IEEE International Symposium on
Information Theory (ISIT'05), p.~1740 (2005). 














\end{thebibliography}
\end{document}